\documentclass[aps,preprint]{revtex4}
\usepackage{amsfonts}
\usepackage{amsmath}
\usepackage{amssymb}
\input{tcilatex}

\begin{document}

\title{Strange nonchaotic attractors in noise driven systems}
\author{Xingang Wang$^{1}$, Meng Zhan$^{1}$, C.-H. Lai$^{2}$, and Ying-Cheng
Lai$^{3}$}
\affiliation{$^{1}$Temasek Laboratories, National University of Singapore, 10 Kent Ridge
Crescent, Singapore 119260 \\
$^{2}$Department of Physics, National University of Singapore, Singapore
117542 \\
$^{3}$Department of Mathematics, Center for Systems Science and Engineering
Research, Arizona State University, Tempe, Arizona 85287 }

\begin{abstract}
Strange nonchaotic attractors (SNAs) in noise driven systems are
investigated. Before the transition to chaos, due to the effect of noise, a
typical trajectory will wander between the periodic attractor and its nearby
chaotic saddle in an intermittent way, forms a strange attractor gradually.
The existence of SNAs is confirmed by simulation results of various critera
both in map and continuous systems. Dimension transition is found and
intermittent behavior is studied by peoperties of local Lyapunov exponent.
The universality and generalization of this kind of SNAs are discussed and
common features are concluded.\newline
\end{abstract}

\maketitle

Since the pioneer work of Grebogi \textit{et al}. in 1984 \cite{grebogi},
strange nonchaotic attractors (SNAs) have been an important topic in
nonlinear dynamics and studied widely both in theory and experiment \cite%
{venkatesan,sna-experiment}. Here the word "strange" refers to the
complicated geometry of the attractor, and the word "nonchaotic" refers to
the no sensitive dependence on initial conditions. Therefore SNAs\ are
denoted by systems which own fractal structure but with nonpositive Lyapunov
exponent. SNAs not only play an important role in our study of transitions
to chaos, but also a general phenomenon in physically relevant situations
and related potential applications are also expected \cite%
{phys-relevant,secure-coms,ott}. While the existence of SNAs is firmly
established, a question that remains interesting is the mechanisms and
routes through which SNAs are created. During the past years, five main
mechanisms or scenarios for the creation of SNAs have been advanced: the
birth of SNAs\ through collision between a period-doubled torus and its
unstable parent \cite{route1}; the collision between a stable torus and an
unstable one at a dense set of points \cite{route3}; the fractalization of a
torus that the increasing wrinkling of tori leads to the appearance of SNAs
without any interaction with a nearby unstable periodic orbit \cite{route2};
the loss of transverse stability of a torus \cite{tolga}; the appearance of
SNAs through type-I\ intermittency or type-III intermittency \cite{route5}.

With our understanding, there are two features which are common for former
studied SNAs: first, SNAs are found typically in qusiperiodically driven
systems; second, the routes and mechanisms mentioned above are not general
in the sense that they are model dependent, namely, has not a universal way
which can be used to create SNAs in most models. Qusiperiod stands a middle
position between period and random in that, on one hand, its spectrum is
discrete which is similar to the periodic signal, and on the other hand,
qusiperiodic signal never repeats itself in time series just\ like the
random signal. Now we know that period can't induce SNAs, but an interesting
question is whether SNAs can be induced by noise? if this is the situation,
how does it works? this is still an open question which has never been
studied before. In comparison with the qusiperiodic forcing, noise is more
general and almost unavoidable in real systems, thus the study of noise
induced SNAs not only will help our further understanding of some basic
problems in nonlinear dynamics, but has a direct connection with experiments
and applications as well. Meanwhile, we noticed that in studies of noise
induced synchronization and noise induced transitions to chaos \cite%
{noisefractal,lai}, the trajectory often show some kind of intermittency and
dimension has a dramatic change at some critical noise intensity, we also
wish the study of SNAs in noise driven systems can characterize some
properties of these processes and explore some underlying mechanisms behind.
Moreover, how to construct a SNA in a general chaotic system is still a
challenge and an open question \cite{shuai}, by this study, we also wish to
find a more general route for SNAs creation.\ 

Our motivation comes from the recent works of Liu and Lai \cite{lai}, where
the transition to chaos in noise driven systems is investigated. One
important result in their work is that noise can induce unstable dimension
variability in a general sense. When noise intensity $D$ is below the
critical value $D_{c}$, a random initial condition leads to a trajectory
confined in the vicinity of the periodic attractor except few transient
chaos initially. When $D>D_{c}$, dramatic change appears where a typical
trajectory doesn't be confined within the vicinity of the periodic
attractor, it will wander between the periodic attractor and its nearby
chaotic saddle, by this way a connection of unstable dimensions between two
distinct sets is established. By studying the asymptotic behavior of the
largest Lyapunov exponent $\lambda $, a power law relation is found between $%
\lambda $ and $D$ for $D\gtrsim D_{c}$, this relation is also valid for $D$
versus\ the unstable dimension variability and the frequency of visit to
chaotic saddle \cite{lai}. Things we interest here is, as the noise
intensity exceeds $D_{c}$, although $\lambda $ increases linearly, its still
possible that $\lambda $ is nonpositive within some range in the sense of
asymptotic behavior, at the same time, a typical trajectory will wander
randomly between two distinct dimension sets, the period state of $d=0$ and
the chaotic saddle of $d>0$ ($d$ represents the dimension), constructs a
fractal dimension gradually. By definiting $D_{0}$ the critical noise
intensity where $\lambda $\ exceeds $0$, SNAs are expected to appear within
range $D_{c}<D<D_{0}$. The main purpose of this work is just to analyze and
characterize SNAs in this range, and try to find a more general way for SNAs
creation.

As the first example, we study SNAs in noise driven logistic map with
function 
\begin{equation}
x_{n+1}=ax_{n}(1-x_{n})+D\xi _{n},  \label{logistic}
\end{equation}%
where $D$ represents the noise intensity and $\xi _{n}$ is the Gaussian
random variable of zero mean and unit variance. For $a=3.8008$ and $D=0$,
system stays in a period $8$ window with the largest Lyapunov exponent $%
\lambda _{p}\approx -0.127<0$. As $D$ exceeds the critical value $%
D_{c}\approx 10^{-5.08}$, $\lambda >\lambda _{p}$ and increases with a power
law scaling, $\lambda -\lambda _{p}\sim (D-D_{c})^{-\alpha }$, with $\alpha $
a system dependent parameter. Further increasing $D$, when $D>D_{0}\approx
10^{-4.95}$, the largest Lyapunov exponent changes to positive and system
becames chaos. The process that $\lambda $ changes with $D$ is plotted in
Fig. 1(a), the range $D_{c}<D<D_{0}$ where $\lambda _{p}<\lambda <0$ is
shown. As we have analyzed above, in this range, due to the effect of noise,
a typical trajectory will stays partially in the periodic attractor and
partially in the chaotic saddle, we present this intermittent behavior in
Fig. 1(b)\ by plotting variable $x_{n}$ versus iteration time $n$ for a
random chosen initial condition. It can be found that this time the
trajectory doesn't be confined within the vicinity of periodic attractor,
but in some iteration intervals, its trajectory cruises the whole phase
space with range $(0,1)$, this made the dimension no longer zero, but some
value between $0$, the dimension of periodic attractor, and $1$, the
dimension of full developed chaotic attractor. Here we employ the
information dimesion, $d=\underset{\varepsilon \rightarrow 0}{\lim }\frac{I}{%
\ln \varepsilon }=\underset{\varepsilon \rightarrow 0}{\lim }%
[\dsum_{i=1,N}\mu _{i}\ln \mu _{i}]/\ln \varepsilon $, to depict the fractal
dimension involved in this intermittent process, $I\ $represents the
distribution of trajectory in the phase space and dimension is estimated by
calculating the slope between $I$ and box size $\varepsilon $. For the same
noise intensity used in Fig. 1(b), we plotted the relation between $I$ and $%
\varepsilon $ in Fig. 1(c), the related information dimension is estimated
to be $d\approx 0.34$. Together considering the largest Lyapunov exponent $%
\lambda \approx -0.05$, SNA is preliminary found. In order to investigate
the relation between noise intensity $D$ and fractal dimension $d$, we plot
Fig. 1(d). A critical transition of dimension can be found at $D_{c}$. For $%
D<D_{c}$, $d$ increases linearly as $D$ increases, this is by no means than
just the effect of noise, because during our estimation of the fractal
dimension we employ the same range of box size $\varepsilon $, while the
periodic attractor becomes thicker as $D$ increases. But when $D\gtrsim D_{c}
$, the fractal dimension has a distinct change in that it increases more
rapidly (the scaling between $D$ and $d$ near $D_{c}$ can be roughly
estimated to be with a power law scaling and with the same exponent $\alpha $
as the relation between $D$ and $\lambda $ near $D_{c}$). The explanation of
this transition as follows, for $D<D_{c}$, although occasionally there can
be trajectory points be "kicked out" of the periodic attractor and stay a
finite time in the chaotic saddle due to some sudden large noise forcing,
this probability is small enough and doesn't effect the dimension in
comparison with the increase of noise intensity. But for $D>D_{c}$, the
dimension induced by unstable dimension variability becomes much strong than
effect of the noise, and the fractal dimension now is dominated by the
behavior of intermittency, more specifically, by temporal trajectories stay
in the chaotic saddle. At the same time, as $D$ increases, the frequency
that a typical trajectory stays in the chaotic saddle also increases in a
power law relation \cite{lai}, as a result, $d$ increases more rapidly after 
$D_{c}$ and a transition of dimension at this point is formed. This
dimension transition shows us a new picture of the critical behavior near $%
D_{c}$, and may help our understanding of $D_{c}$ from another different
aspect,

In order to investigate the behavior of SNAs in more detail, except the two
basic criteria of the nonpositive\ largest Lyapunov exponent and the fractal
dimension, a host of other properties have also been used to characterize
SNAs, such as properties of frequency spectrum and singular-continuous
spectrum, properties of local Lyapunov exponent (LLE), and so on \cite%
{venkatesan}. It is well known that the frequency spectrum of SNAs admits a
power law relation $N(\sigma )\sim \sigma ^{-\kappa }$ with $1<\kappa <2$,
where the spectral distribution function $N(\sigma )$ is defined as the
number of peaks in the Fourier power spectrum larger than some value $\sigma 
$. In Fig. 2(a)\ and (b)\ we plot the power spectrum and the relation
between $N$ and $\sigma $ for trajectory of Fig. 1(b), respectively. The
slope shown in Fig. 2(b) is estimated to be $\kappa \approx $ $0.86$.
Another criterion for SNAs verifying is to analyze its singular-continuous
spectrum, by calculating Fourier transform $X(\Omega
,T)=\sum_{n=1}^{T}x_{n}e^{i2\pi n\Omega }$ with $x_{n}$ the time series of
trajectory and $T$ the total time. For a proper frequency $\Omega $, it was
demonstrated that for SNAs the power scaling $\left\vert X(\Omega
,T)\right\vert ^{2}\sim T^{\beta }$ is hold, and the exponent $\beta \ $is
between $1$ and $2$ which represents a persistent motion (drift) and a
random motion (Brownian motion), respectively. Also, the path for $(\func{Re}%
X,\func{Im}X)$ is expected to be fractal and self-similar \cite%
{Pikovsky-singular,tolga}. In our study, we use the same parameters as Fig.
1(b)\ and set $\Omega =(\sqrt{5}-1)/2$, the golden number, plot $\left\vert
X(\Omega ,T)\right\vert ^{2}$ versus $T$ in Fig. 2(c)\ and $\func{Re}X$
versus $\func{Im}X$ in Fig. 2(d). The slope in Fig. 2(c)\ is estimated to be 
$\beta \approx 1.5$ and the path $(\func{Re}X,\func{Im}X)$ apparently
exhibits a fractal and self-similar structure. These results again confirm
the attractor in Fig. 1(b)\ is strange nonchaotic.

The largest Lyapunov exponent only give the asymptotic value which
represents the average rate of separation of nearby trajectories, it can't
give the temporal information and explore the internal dynamics. To overcome
this shortcoming and also in order to detect the timely behavior of SNAs,
local or finite-time Lyapunov exponent $\lambda _{i}(t)$ was defined and
used for SNAs studies in former works \cite{venkatesan,route5}. The
definition of $\lambda _{i}(t)$ is similar to the largest Lyapunov exponent
except that it is computed over a finite-time interval, $t$, the subscript $i
$ indexes the segment in which this exponent is evaluated. Depending on the
difference of route for SNAs creation, the behavior of LLE is also different
from each other, this property is also regarded as one of the important
criteria for SNAs classification \cite{venkatesan}. The utilization of LLE
in our model is straightforward, when trajectory stays in the vicinity of
the periodic attractor, its behavior is similar to the original periodic
state and LLE is negative, when trajectory stays in the vicinity of the
chaotic saddle, on the contrary, the LLE will be positive. Another advantage
for using LLE exists in its utilization in intermittency studies, because
system is driven by noise, the return position of points escaping from
chaotic saddle will locate to one of periodic states by random, this make it
difficult to quantify the distance that trajectory leaves from the periodic
attractor, but this problem doesn't exist for LLE in the sense that LLE only
consider which sets, period or chaotic saddle, the local trajectory is
stays, regardless which periodic state it locate.\ With $D=10^{-5}$, in Fig.
3(a)\ we plot $\lambda _{i}(t)$ versus segment $i$ for $t=1000$, random
bursting of LLE can be found and typical behavior of intermittency is shown.
In order to explore its temporal and amplitude properties, we plot the
laminar-phase and possibility distributions of $\lambda _{i}$ in Fig. 3(b)
and (c),\ respectively. Significant positive tail, as shown in Fig. 3(c),
decays slowly as a function of $\lambda _{i}$, strongly indicates the
distinction of intermittency and other routes of SNAs creation \cite{route5}%
. A direct consequence of this distinction is the variance of $\lambda _{i}$
increases drastically after the transition point at $D_{c}$, as we plot in
Fig. 3(d). These observations are also consistent with the intermittent
trajectory ( Fig. 1(b)) and the transition of dimension (Fig. 1(d)). The
exponential law of laminar-phase in Fig. 3(c) indicates the happening of
symmetry breaking near $D_{c}$ \cite{tolga,lai}, similar to the type-III
intermittency defined in Ref. \cite{route5}.

We choose Duffing equation as our second model so as to investigate noise
induced SNAs in continuous dynamical systems. As a typical nonautonomous
system, Duffing equation owns a zero Lyapunov exponent which associated with
the time axis, different to autonomous systems (like Lorenz, Rossler and
models studied in Ref. \cite{lai}), the neutral dimension is always there
regardless the noise intensity). Meanwhile, after noise intensity exceeds
the critical value $D_{c}$, the nontrival largest Lyapunov exponent $\lambda 
$ also increases with a power law\ scaling and changes to positive at $D_{0}$
\cite{lai}, thus for Duffing model SNAs there also exists range $%
D_{c}<D<D_{0}$ where SNAs are expected ($D_{c}$ and $D_{0}$ have the similar
definitions as logistic map but for Duffing equation here). The functions we
will study as follows,%
\begin{align}
\overset{.}{x}& =y+D\xi ,  \notag \\
\overset{.}{y}& =-hy+(1+A\cos t)x-x^{3}+D\xi ^{^{\prime }},  \label{duffing}
\end{align}%
with $\xi $ and $\xi ^{^{\prime }}$ two independent Gaussian random
variables of zero mean and unit variance, and $D$ the noise intensity. For $%
h=0.1$ and $A=0.12$, we plot $\lambda $ versus $D$ in Fig. 4(a), there two
critical noise intensities, $D_{c}$ and $D_{0}$, can be found and SNAs are
expected within range $10^{-1.65}\approx D_{c}<D<D_{0}\approx 10^{-1.3}$.
When $D=0$, the trajectory is located in periodic $4$ window and $\lambda
_{p}\approx -0.047$. For $D=10^{-1.4}$, $\lambda \ \approx -0.03$, the time
series of variable $y$ is plotted in Fig. 4(b) and intermittency is shown.
Fig. 4(c)\ shows the $(x,y)$ projection of a trajectory of $10000$
iterations (after $5000$ preiterations) on the stroboscopic surface of
section defined by $t^{\prime }=2n\pi (n=1,2,....)$ for the same parameter
of Fig. 4(b), geometric shape of the attractor appears to be strange and
related information dimension is estimated to be $d\approx 1.3$. In Fig.
4(d)\ we plot the relation between noise intensity $D$ and information
dimension $d$, again, we can find a transition near $D_{c}$ and $d$
increases rapidly once noise intensity exceeds this critical value. For
further study of other properties associated with SNAs, in Fig. 5(a) we plot
the singular-continuous spectrum, the subplot in Fig. 5(a) is the path of $(%
\func{Re}X,\func{Im}X)$ with $X$ the power amplitude of Fourier transform,
the slope, $\beta \approx 1.64$, and fractal structure again confirm the
existence of SNA. In order to investigate its intermittent behavior in
detail, we plot local Lyapunov exponent $\lambda _{i}(t)$ versus $i$ with $%
t=200$ in Fig. 5(b), laminar-phase in Fig. 5(b) and possibility
distributions in Fig.5 (d), respectively. All these results indicate that
this is typically an intermittent behavior and system undergoes a kind of
symmetry breaking. In our simulations, we also test the variance of $\lambda
_{i}$ for Duffing model, just as we find for logistic map, a sudden
transition appears near $D_{c}$ also.

Since the situation where two coexisting dynamical invariant sets with
distinct unstable dimensions can linked by noise can occur in any periodic
window and states where periodic attractor and isolated saddle periodic
orbits coexist, and also because noise is ubiquitous in real systems, the
phenomenon of SNAs induced by noise is expected to be fairly common and
general. At the same time, SNAs not only can be found in transitions from
period to chaos, but also can be found in transitions from chaos to period
and in noisy synchronization systems. For example, for the model used in
Ref. \cite{chlai} with noise level $\sigma ^{2}=0.3$ (details about function
and parameters please refer to the same reference), the largest Lyapunov
exponent $\lambda \approx -0.3$, while the related information dimension is
estimated with $d\approx 1.55$, a behavior of SNA. Moreover, the above
argument can be extended to higher-dimensional systems directly \cite{lai}.
As we have mentioned in the first paragraph, its still remains an open
question to construct a SNA in any chaotic system, by this study we find
that the mechanism of noise induced SNAs is model independent and universal,
so we think this is a more generic method for SNA creation and can be
utilized both in theory and experiment. Meanwhile, this kind of SNAs are not
only typical, but also robust as well \cite{kim}, since the system itself is
under the perturbation of noise. We have also tried perturbations in other
variables and parameters in models studied in this paper, the existence of
SNAs is indeed robust within a wide range. To conclude, we summarize the
common features for noise induced SNAs: (i) It is expected both in map
system and nonautonomous continuous system. For autonomous continuous system
the neutral dimension will be broken and the zero Lyapunov exponent changes
to positive after $D_{c}$, thus SNAs maybe can't be found. (ii) This is
typically an intermittent behavior and associates with some kind of symmetry
breaking. (iii) Both the dimension and the variance of local Lyapunov
exponent have a transition at $D_{c}$, which reflect the change of dynamical
structure from other points of view. (iv) The mechanism is general and the
phenomenon is robust.

We wish this work can help our comprehension in studies like transitions to
chaos, noise induced synchronization \cite{chlai}, particle floating on a
moving fluid \cite{ott}, chaos control in noisy systems \cite{control}, and
so on. Also, we wish the phenomenon of noise induced SNAs can be observed in
experiments in the near future.

Captions of Figures

Fig. 1 For noise driven logistic map. (a) The largest Lyapunov exponent $%
\lambda $ versus noise intensity $D$, SNA is expected within range $%
10^{-5.08}\approx D_{c}<D<D_{0}\approx 10^{-4.95}$. (b) $D=10^{-5.0}$, $%
\lambda \approx -0.05$, intermittent behavior of trajectory, and (c), the
related fractal information dimension $d\approx 0.34$. (d) Fractal dimension 
$d$ versus $D$, a transition exists near $D_{c}$.

Fig. 2 Analysis of trajectory shown in Fig. 1(b). (a) Power spectrum
amplitude $A$ versus frequency $f$, and (b) the related spectrum
distribution with slope $\kappa \approx 0.8$. (c)\ Singular-continuous
spectrum $\left\vert X\right\vert ^{2}$ versus time with slop $\beta \approx
1.5$, and (d) fractal and self-similar path of $(\func{Re}(X),\func{Im}(X))$.

Fig. 3 Parameters same to Fig. 1(b). (a) Intermittency behavior of local
Lypunov exponent, (b) the power law laminar-phase of (a). (c) Possibility
distribution of $\lambda _{i}$ shown in (a), and (d) variance of $\lambda
_{i}$ versus noise.

Fig. 4 For noise driven Duffing equation. (a) The largest Lyapunov exponent $%
\lambda $ versus noise intensity $D$, SNA is expected within range $%
10^{-1.65}\approx D_{c}<D<D_{0}\approx 10^{-1.3}$. (b) $D=10^{-1.4}$, $%
\lambda \approx -0.047$, the intermittency of variable $y$, and (c)
projection of trajectory on the stroboscopic surface section, the related
fractal information dimension $d\approx 1.3$. (d) Fractal dimension $d$
versus $D$, a transition exists near $D_{c}$.

Fig. 5 Parameters same to Fig. 4(b). (a) singular-continuous spectrum $%
\left\vert X\right\vert ^{2}$ versus time with slope $\beta \approx 1.64$.
(b) Intermittency behavior of local Lypunov exponent, and (c) the power law
laminar-phase of (b). (d) Possibility distribution of $\lambda _{i}$ shown
in (a).

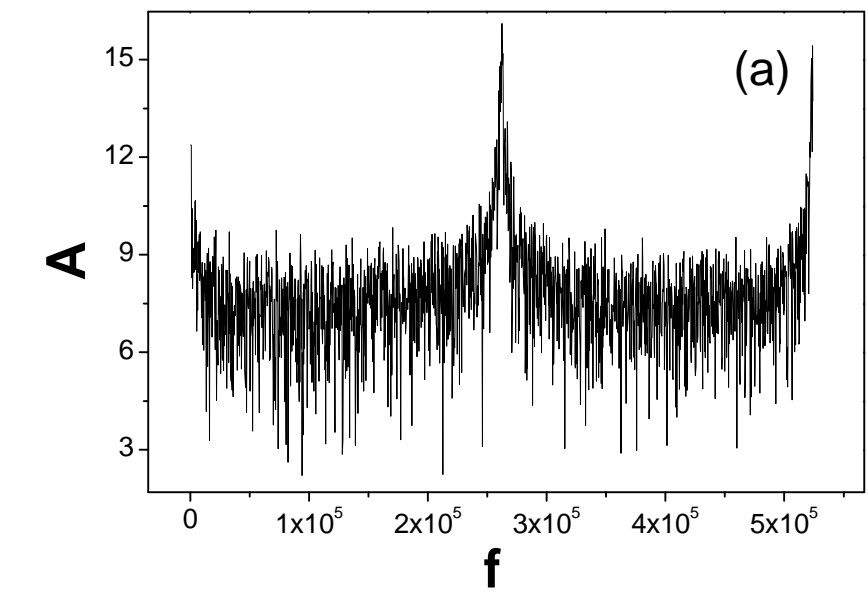
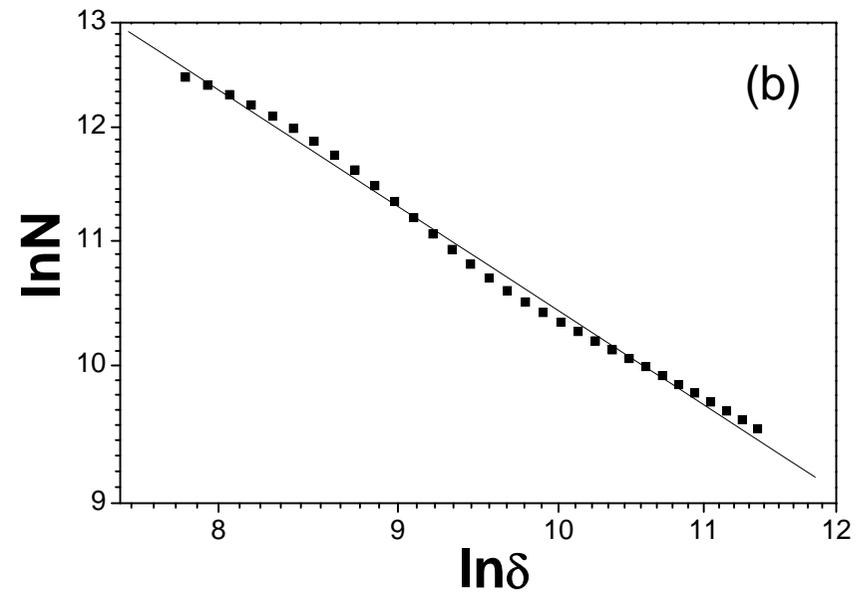
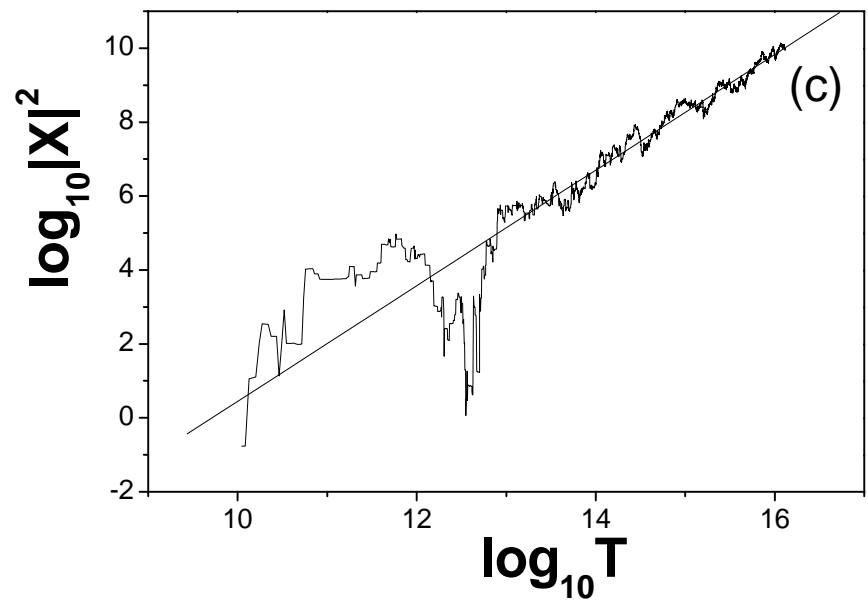
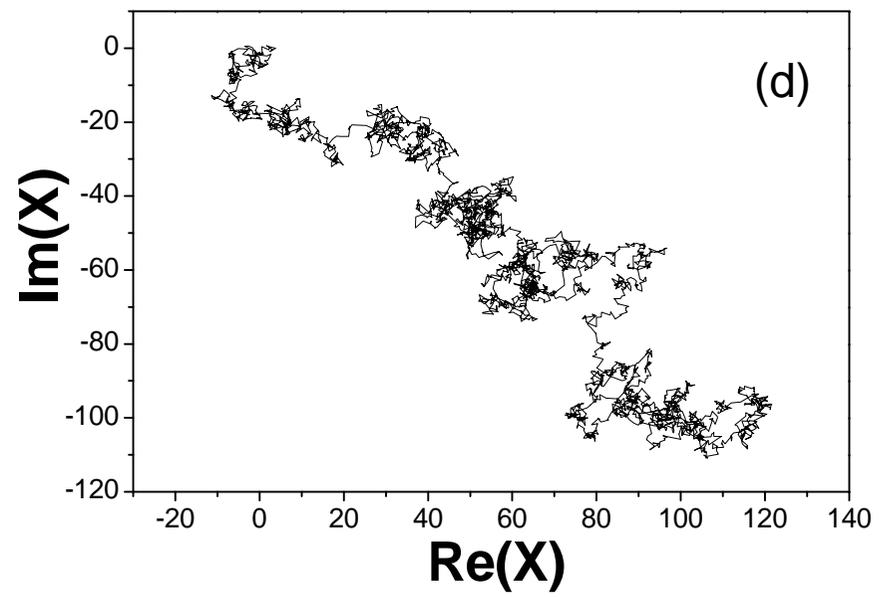
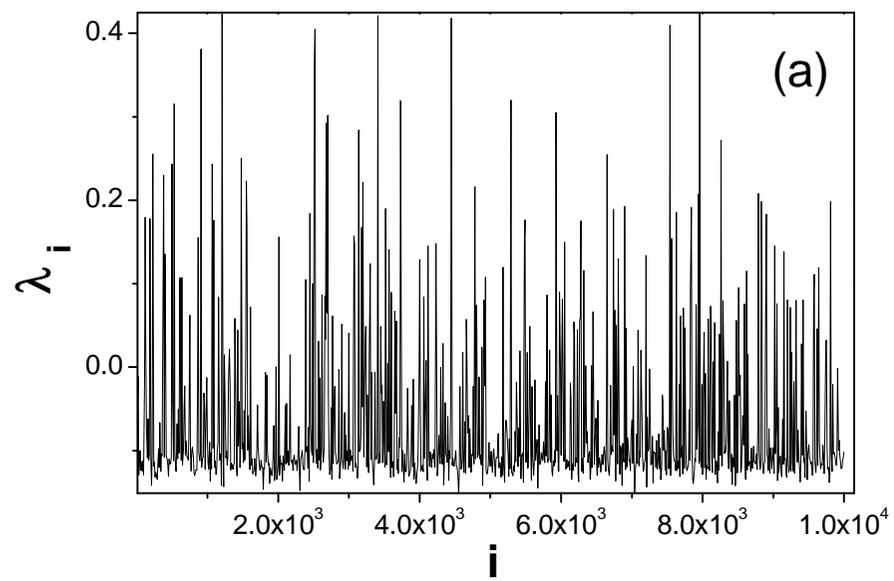
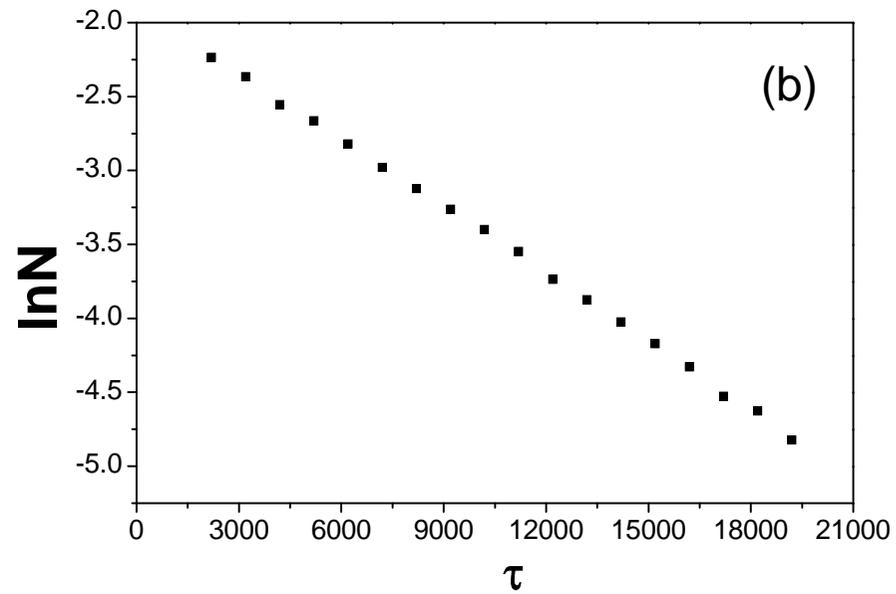
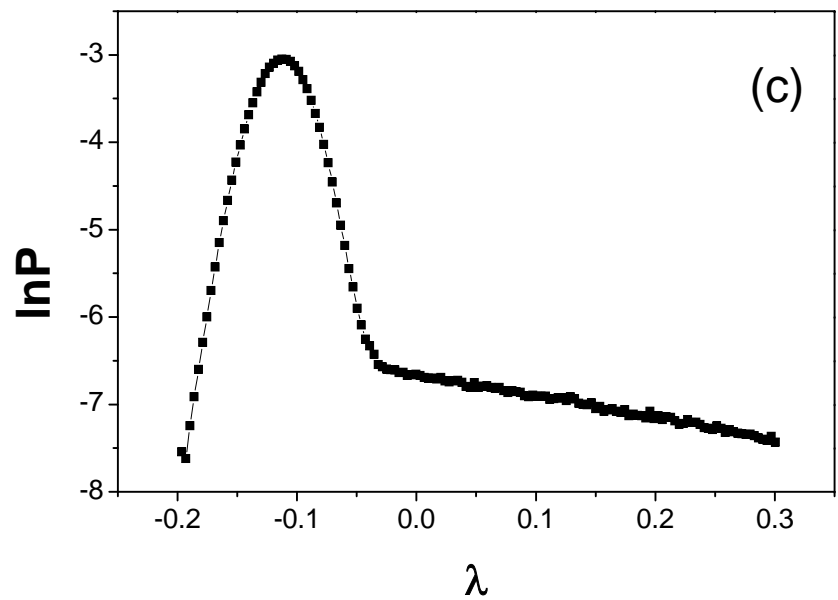
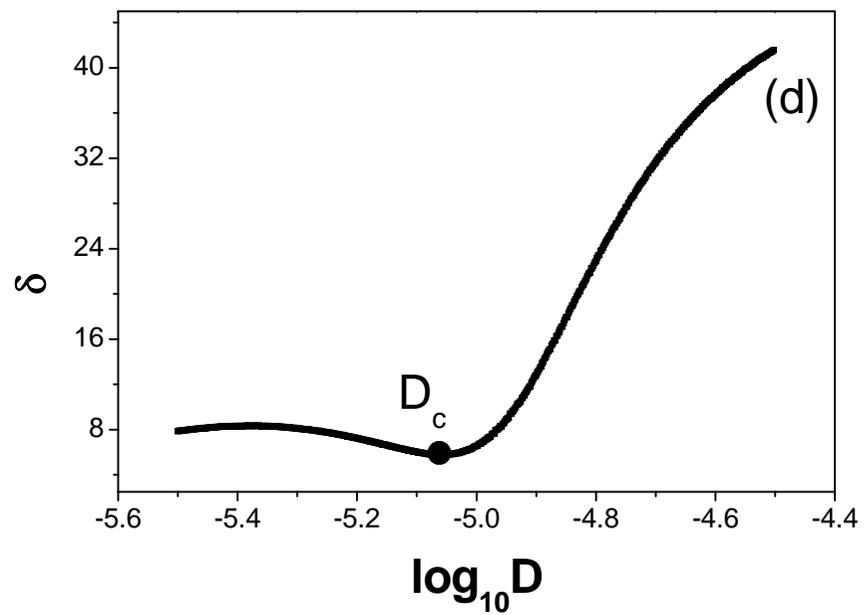
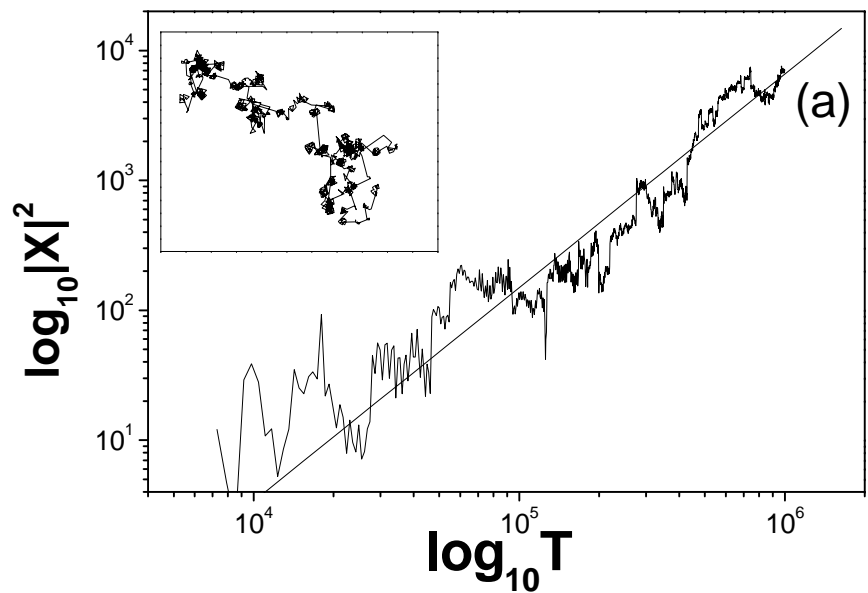
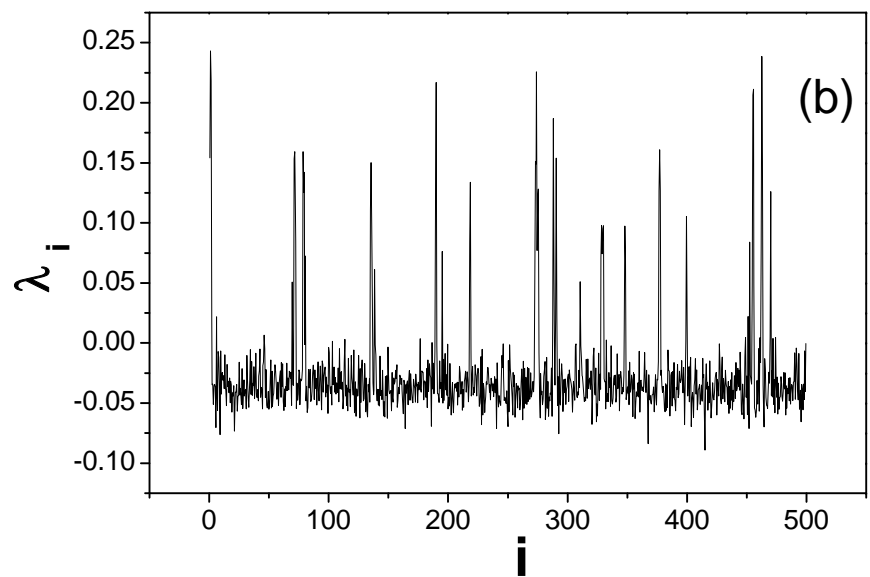
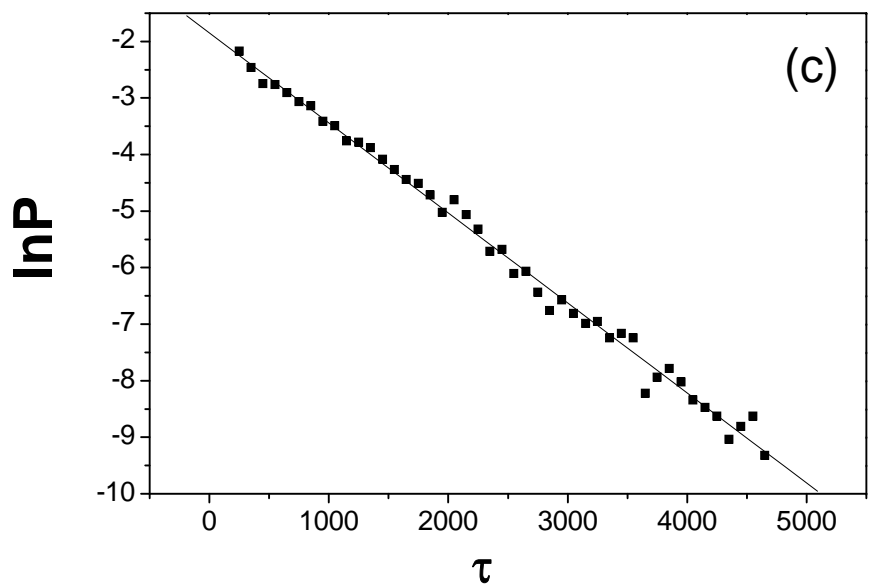
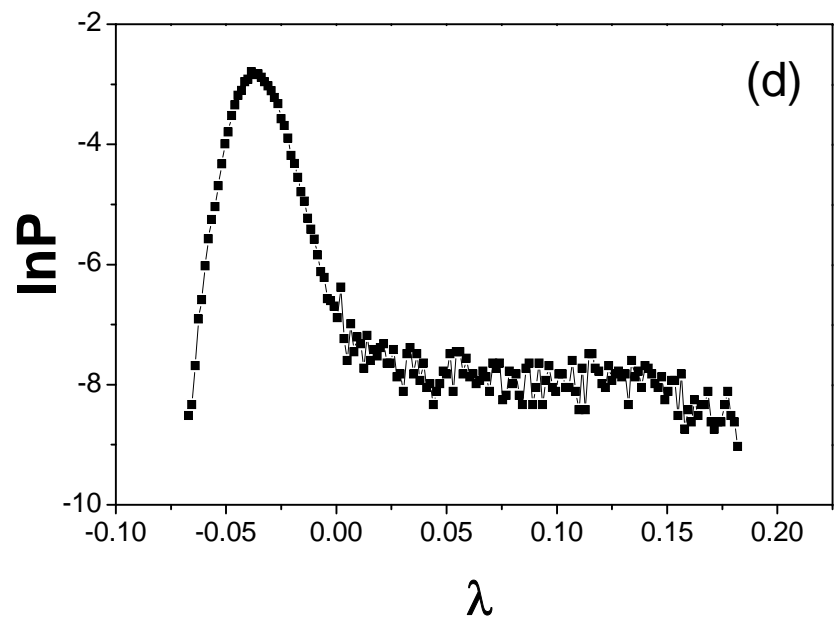
\end{document}